\documentclass[notitlepage,a4paper,aps,prd,superscriptaddress,nofootinbib,groupedaddress,twocolumn]{revtex4}

\usepackage{enumerate}
\usepackage{amsmath}
\usepackage{amsfonts}
\usepackage{amssymb}
\usepackage[utf8]{inputenc}
\usepackage[T1]{fontenc}
\usepackage{mathtools}
\usepackage{wasysym}
\usepackage{accents}
\usepackage{graphicx}
\usepackage{subcaption}
\usepackage[english]{babel}
\usepackage[svgnames]{xcolor}
\usepackage[colorlinks=true,citecolor=blue]{hyperref}
\usepackage{url}
\usepackage{float}
\usepackage{mhchem}
\usepackage{bbold}

\newcommand{\JP}{
\affiliation{Physics Department, Federal University of Para\'iba, Caixa Postal 5008, 58059-900, Jo\~ao Pessoa, PB, Brazil}
}

\newcommand{\Areia}{
\affiliation{Department of Chemistry and Physics, Federal University of Para\'iba, Rodovia BR 079 - km 12, 58397-000 Areia-PB, Brazil}
}

\newcommand{\CG}{
\affiliation{Physics Department, Federal University of Campina Grande, Caixa Postal 10071, 58429-900, Campina Grande-PB, Brazil}
}

\newcommand{\Lavras}{
\affiliation{Department of Physics, Federal University of Lavras, Caixa Postal 3037, 37200-900 Lavras-MG, Brazil}
}

\begin{document}

\title{LIV-Decoherence on Gravitational Cat States}

\author{Iarley P. Lobo}
\email{iarley.lobo@academico.ufpb.b,\\ lobofisica@gmail.comr}
\Areia
\CG

\author{Kelvin Sampaio}\email{kelvin.costa.rocha.sampaio@academico.ufpb.br,\\ sampaiofisica@gmail.com}
\JP

\author{Gislaine Varão}\email{gvs@academico.ufpb.br,\\ gislainevarao@gmail.com}
\JP

\author{Moises Rojas}\email{moises.leyva@ufla.br}
\Lavras

\author{Valdir B. Bezerra}\email{valdir@fisica.ufpb.br}
\JP

\begin{abstract}
Inspired by approaches based on the stochastic generalized uncertainty principle, we propose a Lindblad equation derived from the quantization of a stochastic modified dispersion relation in a Lorentz Invariance Violation (LIV) scenario. This framework enables us to investigate decoherence effects in a system of particles exhibiting gravitationally induced entanglement. We analyze the impact of LIV on entanglement (quantified by concurrence) considering systematic and stochastic effects.
\end{abstract}

\maketitle

%%%%%%%%%%%%%%%%%%%%%%%%%%%%%%%%%%%%%%%%%%%%%%%%%%%%%%%
%%%%%%%%%%%%%%%%%%%%%%%%%%%%%%%%%%%%%%%%%%%%%%%%%%%%%%%
%%%%%%%%%%%%%%%%%%%%%%%%%%%%%%%%%%%%%%%%%%%%%%%%%%%%%%%

\section{Introduction}

Recent years have seen intense research on quantum gravity phenomenology, particularly through high-energy astroparticles that probe spacetime's quantum nature~\cite{Addazi:2021xuf}. Many quantum gravity approaches predict Planck-scale deviations from relativistic kinematics, making high-energy particles ideal probes. However, different theoretical approaches suggest quantum gravity effects may also manifest in the non-relativistic regime through ultraviolet/infrared (UV/IR) mixing~\cite{Amelino-Camelia:2008aez}.

This UV/IR mixing has motivated studies of modified uncertainty principles with minimal length scales~\cite{Petruzziello:2020wkd}, deformed Galilean algebras~\cite{Arzano:2014cya,Arzano:2022nlo,Lobo:2025vqt}, and non-relativistic dispersion relations~\cite{Wagner:2023fmb}. Such effects may appear either as systematic deformations, governed by a fixed quantum gravity scale, or stochastic fluctuations, where the quantum gravity governing parameter is allowed to fluctuate over an average value, both emerging from different quantum gravity approaches~\cite{Amelino-Camelia:2008aez}. While systematic effects have been extensively studied~\cite{Addazi:2021xuf,Amelino-Camelia:2009wvc}, stochastic modifications, although also relevant in different areas of high energy phenomenology \cite{Vasileiou:2015wja} remain less explored, particularly in quantum mechanical systems.

Recent work~\cite{Petruzziello:2020wkd} demonstrated that stochastic modifications to the Schrödinger equation induce fundamental decoherence. Although suppressed by the Planck scale ($\sim 10^{-35}$ m), these universal effects should impact gravitational quantum systems. A prime example is the gravitational cat state (gravcat) - massive particles in superposition that become entangled through their gravitational interaction~\cite{Bose:2017nin,Marletto:2017kzi}. Such systems have driven significant advances in quantum gravity phenomenology~\cite{Rijavec:2020qxd,Rojas:2021eyg}, relativistic quantum information~\cite{Galley:2023byb}, and experimental techniques~\cite{Westphal:2020okx}.

In this work, we investigate how stochastic quantum gravity affects entanglement generation in gravcat states. We develop a first-order stochastic Schrödinger equation (stronger than previous results~\cite{Petruzziello:2020wkd}) and derive the corresponding Lindblad master equation (Section~\ref{sec:2}). After reviewing gravcat properties (Section~\ref{sec:3}), we analyze decoherence effects on entanglement (Section~\ref{sec:4}), discussing implications for realistic experimental parameters. We conclude in section \ref{sec:6}.

%%%%%%%%%%%%%%%%%%%%%%%%%%%%%%%%%%%%%%%%%%%%%%%%%%%%%%%%%%%%%%%%%%%%%%%%%%%%%%%%%%%%%%%%%%%%%%%%%%%%%%%%%%%%%%%%%%%%

\section{Master Equation from Stochastic Dispersion Relation}\label{sec:2}

From a phenomenological perspective, one of the most promising avenues of research in quantum gravity involves the possibility that the quantum nature of spacetime leaves imprints on the kinematical properties of particles. These departures from classical behavior couple with the energy and momentum of the test particles themselves and are suppressed by the quantum gravity energy scale. This effect can be modeled through an effective modification of the Hamiltonian. Different quantum gravity models predict distinct ways in which such departures may occur. A phenomenological approach to capture these corrections and examine their signatures involves considering a modified dispersion relation for particles with energy $E$, momentum $p$, and mass $M$ of the form:
\begin{equation}\label{eq:mdr1}
M^2 c^2 = \left( \frac{E}{c} \right)^2 - p^2 - \ell \xi_{n}(Mc)^{n-3} p^{3-n},
\end{equation}
where $\ell^{-1}$ is the quantum garvity scale, which here has dimensions of momentum and we expect it to be of the order of the Planck momentum $E_{\text{P}}/c\approx 6.5\, \text{N.s}$, where $E_{\text{P}}\approx 1.22\times 10^{19}\, \text{GeV}$ is the Planck energy, $c$ is the speed of light, and $\xi_{n}$ are parameters expected to be of order unity. As discussed previously, this expression incorporates general corrections that depend on the energy and momentum of the test particles.

Relativistic astrophysics provides the most common scenario for searching for effects resulting from these corrections, as they are amplified by particle energies. However, recent studies have suggested that such corrections might also leave imprints at nonrelativistic scales, which are currently being constrained with Planck-scale sensitivity. In this case, as discussed in \cite{Wagner:2023fmb}, the most relevant contributions come from corrections proportional to powers of the momentum and mass:
\begin{equation}\label{eq:mdr2}
E = \frac{p^2}{2M} + \frac{\ell}{2M} \xi_n (Mc)^{3-n} p^n.
\end{equation}

Quantum systems have provided the most important window for investigating these infrared effects. By performing a spacetime foliation and either quantizing Eq.~\eqref{eq:mdr2} or calculating the nonrelativistic limit of the Klein-Gordon equation compatible with \eqref{eq:mdr2}, the Schrödinger equation takes the form \cite{Wagner:2023fmb}
\begin{align}\label{eq:hath1}
    \hat{H}&=\hat{K}+e\phi_e+M\phi_g-\xi_n\ell (Mc)^{3-n}(2M)^{\frac{n-2}{2}}\hat{K}^{n/2},\\
    \hat{K}&=-\frac{h^{ij}}{2M}\left(\nabla_i-ieA_i^{e}\right)\left(\nabla_j-ieA_j^{e}\right)\label{eq:khat}
\end{align}
where $\hat{K}$ is the kinetic energy with a minimal coupling with the electromagnetic potential (in this case, we do have crontributions from a gravitomagnetic potential), $\phi_e$ is the electromagnetic scalar potential on a particle of charge $e$, $\phi_g$ is the scalar gravitational potential on a particle of mass $M$ and $h^{ij}$ are the space components of the metric in a $3+1$-decomposition. 

We refer to $\hat{H}_0=\hat{K}+e\phi_e+M\phi_g\doteq\hat{K}+\hat{V}$ as the undeformed Hamiltonian, and $\hat{H}_{\ell,n}=-\xi_n\ell (Mc)^{3-n}(2M)^{\frac{n-2}{2}}\hat{K}^{n/2}$ represents the Hamiltonian correction due to MDR. This Hamiltonian governs the time evolution of the state vector $|\phi\rangle$, and its corresponding pure state density matrix $\varrho=|\psi\rangle\langle\psi|$ follows the Liouville-von Neumann equation
\begin{equation}
   i\hbar\partial_t|\psi\rangle= (\hat{H}_0+\hat{H}_{\ell,n})|\psi\rangle\Rightarrow\partial_t\varrho=-\frac{i}{\hbar}[\hat{H}_0+\hat{H}_{\ell,n},\varrho].
\end{equation}

In order to analyze this equation, we follow the steps described in \cite{Donadi:2024amp}. We define a new Hamiltonian and its stochastic perturbation as
\begin{align}
    \hat{H}_0'&=\hat{H}_0+g(\hat{K})\, ,\\
    \hat{H}_{\ell,n}'&=\hat{H_{\ell,n}}-g(\hat{K})\, ,
\end{align}
where $g(\hat{K})=\langle \hat{H}_{\ell,n}\rangle$ and $\langle\ast\rangle$ means averaging over the stochastic process. This guarantees that $\langle \hat{H}_{\ell,n}'\rangle=0$.

The stochastic nature of $\xi_n$ will induce a master equation describing an open quantum system, as we now demonstrate. Following the approach in \cite{Petruzziello:2020wkd}, we treat the deformed Hamiltonian contribution $\hat{H}_{\ell,n}$ as describing the particle's interaction with gravitational degrees of freedom. We work in the interaction picture by defining the state $|\psi^I\rangle$ as
\begin{equation}
    |\psi\rangle= e^{-i \hat{H}_0' t/\hbar}|\psi^I\rangle.
\end{equation}

This allows us to write the Schrödinger equation as
\begin{equation}
    i\hbar\partial_t|\psi^I\rangle=e^{i\hat{H}_0't/\hbar}\hat{H}_{\ell,n}'e^{-i\hat{H}_0't/\hbar}|\psi^I\rangle\doteq\hat{H}'^I_{\ell,n}|\psi^I\rangle.
\end{equation}

Consequently, the Liouville-von Neumann equation for the density matrix $\varrho^I= |\psi^I\rangle\langle\psi^I|$ in the interaction picture becomes
\begin{equation}\label{eq:rhotilde1}
    \partial_t\varrho^I=-\frac{i}{\hbar}[\hat{H}'^I_{\ell,n},\varrho^I].
\end{equation}

The formal solution to this equation is
\begin{equation}\label{eq:sol1}
    \varrho^I(t)=\varrho^I(0)-\frac{i}{\hbar}\int_0^t[\hat{H}'^I_{\ell,n}(t'),\rho^I(t')]dt'.
\end{equation}

Substituting \eqref{eq:sol1} back into \eqref{eq:rhotilde1} yields
\begin{align}
    \partial_t\varrho^I(t)=-\frac{i}{\hbar}[\hat{H}'^I_0(t),\varrho^I(0)]\nonumber\\
    -\frac{1}{\hbar^2}\int_0^t\left[\hat{H}'^I_0(t),[\hat{H}'^I_0(t'),\varrho^I(t')]\right]dt',
\end{align}
where we employ the Born-Markov approximation to insert $\hat{H}'^I_0(t)$ inside the integral as a function of $t$ only. We treat $\xi_n=\sqrt{t_{QG}}\chi_n(t)$ as Gaussian noise with fixed average and sharp autocorrelation:
\begin{align}
    \langle\xi_n(t)\rangle=1, \quad \langle \chi_n(t),\chi_n(t')\rangle=\delta(t-t'),
\end{align}
where $t_{QG}$ is the quantum gravity timescale, presumably of Planck-time order.

Since the evolution equation fluctuates, we consider the average density matrix $\langle\varrho\rangle\doteq\rho$ to study the quantum system. The average differential equation can be found using the cumulant expansion method, expanding $\rho^I$'s equations of motion in terms of the cumulant of the operator $[\hat{H}_{\ell}'^I,\ast]$ \cite{1974Phy....74..215V}:
\begin{equation}
    \partial_t\rho^I=-\int_0^t\langle\left[\hat{H}'^I_{\ell,n}(t),[\hat{H}'^I_{\ell}(t'),\rho^I(t)]\right]\rangle dt'.
\end{equation}

Assuming $\hat{H}'^I_{\ell,n}(t)$ follows Gaussian white noise statistics yields the evolution equation:
\begin{align}
     \partial_t\rho^I=-\frac{\sigma_n^2}{\hbar^2}t_{QG}\left[(\hat{K}^I)^{n/2},[(\hat{K}^I)^{n/2},\rho^I]\right],
\end{align}
where $\sigma_n=\ell(Mc)^{3-n}(2M)^{\frac{n-2}{2}}$. For the density matrix $\rho=e^{-i\hat{H}_0't/\hbar}\rho^Ie^{i \hat{H}_0't/\hbar}$, we obtain our main result -- a Lindblad equation:
\begin{align}\label{eq:main}
    \partial_t\rho(t)=-\frac{i}{\hbar}\left[\hat{K}+\hat{V}-\sigma_n\hat{K}^{n/2},\rho(t)\right]\nonumber\\
    -\frac{\sigma_n^2}{\hbar^2}t_{QG}\left[\hat{K}^{n/2}(t),\left[\hat{K}^{n/2}(t),\rho(t)\right]\right].
\end{align}

This equation describes an open quantum system experiencing decoherence due to quantum spacetime fluctuations. The interaction strength with the environment is characterized by $\sigma_n$, which depends on the system's mass. For $n=1$, our Lindbladian $\sqrt{\hat{K}}\propto \hat{p}$ resembles that in \cite{Arzano:2022nlo}, though the latter uses momentum components $\hat{p}_i$ as Lindbladians. For $n=2$, the Lindblad term becomes similar to \cite{Breuer:2008rh}, with $\hat{p}^2\propto \hat{K}$ as the Lindbladian. The $n=3$ case introduces a novel $\hat{K}^{3/2}$ term.

To estimate the decoherence timescale, we consider a free nonrelativistic particle with $\hat{H}_0=\hat{K}-\sigma_nK^{n/2}$. In the momentum representation, the density matrix elements $\rho_{p,q}=\langle p|\rho|q\rangle$ (where $|p\rangle$ is a momentum eigenstate) evolve as:
\begin{equation}
    \rho_{pq}(t)=\rho_{pq}(0)\exp\left[-\frac{i}{\hbar}\left(\Delta{\tilde{E}}\right)t\right]\exp\left[-\sigma_n\left(\Delta E^{n/2}\right)^2 t\right].
\end{equation}
where $\tilde{E}=E-\sigma_n E^{n/2}$, $\Delta \tilde{E}=\tilde{E}(p)-\tilde{E}(q)$ and $\left(\Delta E^{n/2}\right)^2=\left(E^{n/2}(p)-E^{n/2}(q)\right)^2$. Off-diagonal elements are suppressed by the $\sigma_n$ damping factor, with decoherence time $\tau_D=1/(\sigma_n\left(\Delta E^{n/2}\right)^2)$. Substituting $\ell=c/E_P$, the decoherence time becomes:
\begin{equation}\label{eq:dec-time1}
    \tau_{D,n}=\frac{2^{2-n}\hbar ^2}{(Mc^2)^{4-n}\left(\Delta E^{n/2}\right)^2}\frac{E_P^2}{t_{QG}}.
\end{equation}

For comparison with relativistic quantum gravity decoherence models, we set $M=E/c^2$ and $\Delta E^{n/2}\sim E^{n/2}$. Taking $t_{QG}=t_P=\hbar/E_P$ yields an $n$-independent decoherence time:
\begin{equation}
    \tau_D \propto \hbar\frac{E_P^3}{E^4},
\end{equation}
sharing the energy dependence of the Ellis-Mohanty-Nanopoulos model \cite{Ellis:1988uk}. Our goal is to examine LIV effects on entanglement in gravitationally interacting massive quantum states.

%%%%%%%%%%%%%%%%%%%%%%%%%%%%%%%%%%%%%%%%%%%%%%%%%%%%%%%%%%%%%%%%%%%%%%%%%%%%%%%%%%%%%%%%%%%%%%%%%%%%%%%%%%%%%%%%%%%%

\section{Gravitational Cat States}\label{sec:3}

\begin{figure}
    \centering
\includegraphics[width=0.6\linewidth]{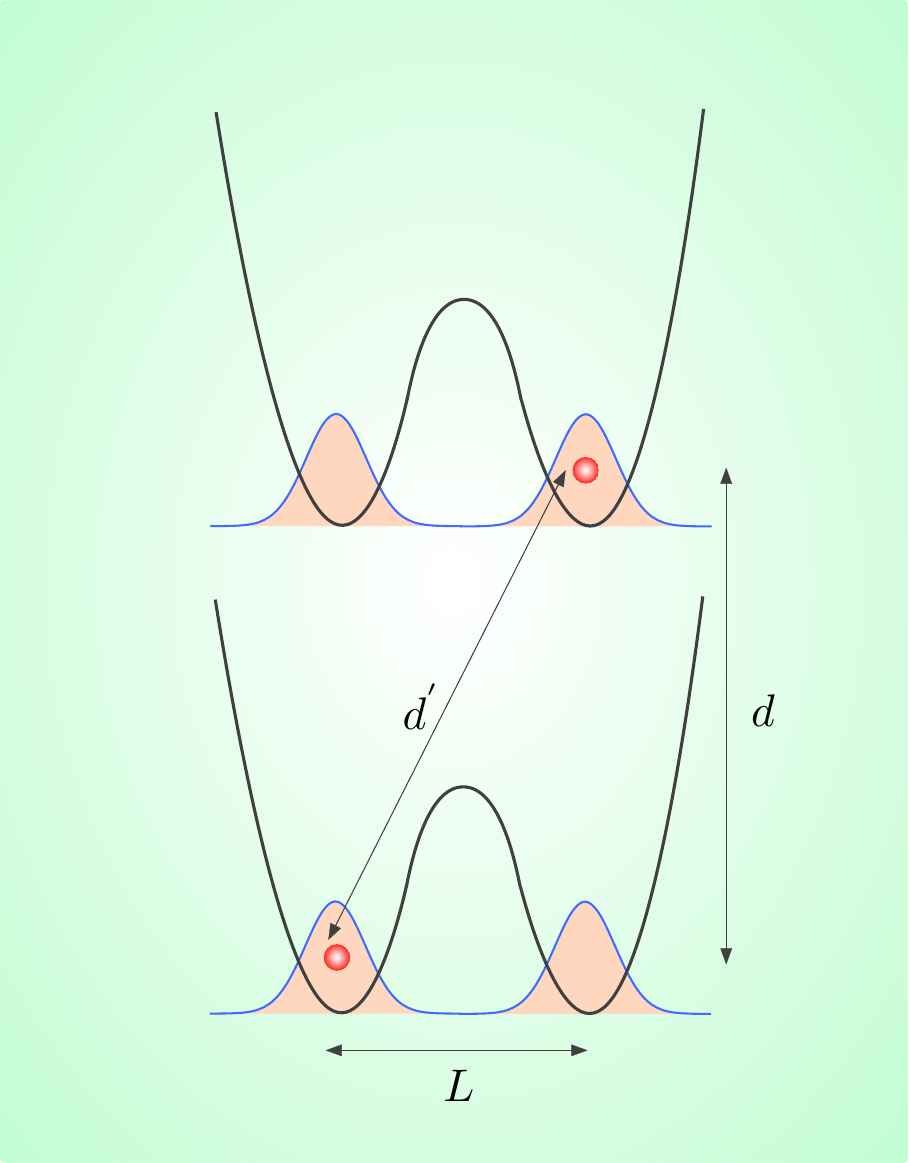}
    \caption{Representation of the gravcat states. The two particles are located in an even double-well potential with minima separated by distance $L$. The interparticle distance is $d$ or $d'$ when the particles occupy the same or different potential minima, respectively.}
    \label{fig:double-potential}
\end{figure}

Consider a system of two particles, each confined in a double-well potential with local minima at positions $x_{\text{min}}=\pm L/2$ that interact gravitationally as in Fig.\ref{fig:double-potential}.  Under the conditions discussed in \cite{Anastopoulos:2020cdp}, where the second excited energy level lies far above the first excited and ground states, we can approximate each particle as a two-level system. The Hamiltonian for a two-level particle is given by the $z$-Pauli matrix $\hat{K}=\frac{\epsilon}{2}\sigma_z$, with eigenstates $|0\rangle$ and $|1\rangle$ corresponding to eigenenergies $\pm \epsilon/2$, where $\epsilon$ represents the energy gap between states. These states define the computational basis.

Since the particles can be well-localized in either the left or right well of the double-well potential, they can also be described by superpositions of left and right position eigenstates, denoted $|\pm\rangle$. These particles are in in a superposition of positions and energies. Due to the uncertainty in the location of the each particle, there is a superposition in the gravitational potential that describes their interaction. We refer to these as gravitational cat (gravcat) states.

The system Hamiltonian is given by \cite{Anastopoulos:2020cdp}:
\begin{align}\label{eq:und-ham}
\hat{H}_0=(E-\Gamma)\mathbb{1}\otimes\mathbb{1}+\frac{\epsilon}{2}(\mathbb{1}\otimes\sigma_z+\sigma_z\otimes \mathbb{1})-\Omega \sigma_x\otimes\sigma_x,
\end{align}
where $E>\epsilon$ is a reference energy shifted by $\pm \epsilon$, $\Gamma=\frac{1}{2}\alpha\left(\frac{1}{d}+\frac{1}{d'}\right)$, $\Omega=\frac{1}{2}\alpha\left(\frac{1}{d}-\frac{1}{d'}\right)$, and $\alpha=Gm^2$. While the constant terms proportional to $E$ and $\Gamma$ do not contribute to Lindblad evolution, we retain them in the Hamiltonian because in the following section we will consider $\hat{K}^{n/2}$ as the Lindbladian, which requires a positive-definite kinetic energy operator.

The Liouville-von Neumann evolution $\partial_t\rho=-i[\hat{H}_0,\rho(t)]$ for the initial state $|\psi_0\rangle=\sin(\theta)|00\rangle+\cos(\theta)|11\rangle $
\begin{align}
\rho(0)=\left(
\begin{array}{cccc}
 \cos^2(\theta) & 0 & 0 & \sin(\theta)\cos(\theta) \\
 0 & 0 & 0 & 0 \\
 0 & 0 & 0 & 0 \\
 \sin(\theta)\cos(\theta) & 0 & 0 & \sin^2(\theta) \\
\end{array}\label{eq:init-cond}
\right)
\end{align}
yields the following non-zero matrix elements:
\begin{widetext}
\begin{align}
    \rho_{11}(t)&=\frac{\Omega^2-\Omega \epsilon \sin(2\theta)+\epsilon^2 \cos(2\theta)+\Omega \cos\left(2t\frac{\sqrt{\Omega^2+\epsilon^2}}{\hbar}\right)(\Omega \cos(2\theta)+\epsilon \sin(2\theta))+\epsilon^2}{2(\Omega^2+\epsilon^2)},\label{eq:sol1}\\
    \rho_{14}(t)&=\frac{1}{2(\Omega^2+\epsilon^2)}\left[\Omega \cos(2\theta)\left(-i\frac{\sqrt{\Omega^2+\epsilon^2}}{\hbar}\sin\left(2t\frac{\sqrt{\Omega^2+\epsilon^2}}{\hbar}\right)+\epsilon \cos\left(2t\frac{\sqrt{\Omega^2+\epsilon^2}}{\hbar}\right)-\epsilon\right)\right.\nonumber\\
    &\left.+\sin(2\theta)\left(\Omega^2-i\epsilon\frac{\sqrt{\Omega^2+\epsilon^2}}{\hbar}\sin\left(2t\frac{\sqrt{\Omega^2+\epsilon^2}}{\hbar}\right)+\epsilon^2 \cos\left(2t\frac{\sqrt{\Omega^2+\epsilon^2}}{\hbar}\right)\right)\right],\\
  \rho_{44}(t)&=\frac{\Omega^2+\Omega \epsilon \sin(2\theta)-\epsilon^2 \cos(2\theta)-\Omega \cos\left(2t\frac{\sqrt{\Omega^2+\epsilon^2}}{\hbar}\right)(\Omega \cos(2\theta)+\epsilon \sin(2\theta))+\epsilon^2}{2(\Omega^2+\epsilon^2)}.\label{eq:sol2}
\end{align}
\end{widetext}
For the initial condition with $\theta=0$ (an unentangled state), the system evolves into an entangled state, as quantified by the concurrence \cite{Wootters:1997id}:
\begin{equation}
     \mathcal{C}(t) = 2 \max \left\{ |\rho_{23}| - \sqrt{\rho_{11} \rho_{44}}, |\rho_{14}| - \sqrt{\rho_{22} \rho_{33}}|, 0 \right\}.\label{eq:conc}
\end{equation}

The characteristic timescale for entanglement generation is:
\begin{equation}\label{eq:ent-time}
    \tau_{E}=\frac{\hbar}{\sqrt{\Omega^2+\epsilon^2}}.
\end{equation} 

This timescale $\tau_E$ determines the oscillation period of the entanglement, while $\Omega$ governs the entanglement strength, as shown in Fig.(\ref{fig:grid}\textcolor{red}{a}).

For an initially unentagled state $\theta=0$, we obtain:
\begin{equation}
 \mathcal{C}_{\theta=0}(t)= 2\frac{\Omega}{\hbar^2}\tau_E^2|\sin(t/\tau_E)|\sqrt{\epsilon^2+\Omega^2\cos^2(t/\tau_E)}.
\end{equation}

Notably, when $\theta=0$ and $\Omega=0$, we find $\mathcal{C}=0$. This observation underlies the concept that particles in position superposition states can develop entanglement mediated by their gravitational interaction.

The next step involves investigating the effects of Lorentz invariance violation on this system.
%%%%%%%%%%%%%%%%%%%%%%%%%%%%%%%%%%%%%%%%%%%%%%%%%%%%%%%%%%%%%%%%%%%%%%%%%%%%%%%%%%%%%%%%%%%%%%%%%%%%%%%%%%%%%%%%%%%%

\section{LIV Effect on Entanglement Between Gravcat States}\label{sec:4}

Consider that the double-well potential of each particle is generated by the interaction of our system with an electromagnetic field. This could be realized, for instance, via the minimal coupling of the system with the magnetic potential in the $y$ direction of the form $\vec{A}=(0,bx^2,0)$ in Cartesian coordinates. In fact, according to the prescription \eqref{eq:khat}, in the position representation, the Hamiltonian (kinetic term with minimal coupling) of this system is of the form 
\begin{equation}
    \hat{K}=\frac{1}{2M}\left[p_z^2+\hat{P}_x^2+e^2b^2\left(\frac{p_y}{eb}-x^2\right)^2\right]\, ,
\end{equation}
where $p_y$ and $p_z$ are the momenta of the particle and $\hat{P}_x$ is the momentum operator in the $x$-direction. Apart from the constant displacement given by $p_z$, the interaction term $e^2b^2(p_y/eb-x^2)^2$ is a double well potential, whose shape can be controlled by $p_y$, $e$ and $b$ and the interwell coupling occurs in the $x$-direction. An example of such system consists in those found from Bose-Einstein condensation, for instance in \cite{PhysRevA.55.4318}, some conditions are given on the potential such that, in each well, the two lowest energy states are closely spaced and well separated from higher energy levels of the potential. Although hard to be realized experimentally due to electrostatic repulsion, in principle, this system is can be formed under certain conditions \cite{Lukin_2021}.

Therefore, assuming that the two-level configuration is generated by a minimal coupling with $\vec{A}$, we can identify the kinetic energy operator as $\hat{K}=E\mathbb{1}\otimes\mathbb{1}+\frac{\epsilon}{2}(\mathbb{1}\otimes\sigma_z+\sigma_z\otimes \mathbb{1})$. From Eq.\eqref{eq:main}, powers of this operator will play the role of Lindbladians in this analysis.

Since $\hat{K}$ is diagonal, we can directly consider its components raised to the power $n/2$, which remain real-valued due to the condition $E>\epsilon$. This ensures the Hermiticity of the kinetic energy operator, which serves as the Lindblad operator in this approach.

Substituting this information into the Lindblad evolution some components of the density matrix evolve in an undeformed way: $\rho _{11}'(t)=-\rho _{44}'(t)=-i \Omega \left(\rho _{14}(t)-\rho _{41}(t)\right)/\hbar,\quad  \rho _{22}'(t)=-\rho _{33}'(t)=-i \Omega \left(\rho _{23}(t)-\rho _{32}(t)\right)/\hbar,\quad \rho _{23}'(t)=-i \Omega \left(\rho _{22}(t)-\rho _{33}(t)\right)/\hbar$, while the LIV effects are restricted to the component
\begin{align}
   \rho _{14}'(t) &= -\frac{A_n^2t_{QG}}{\hbar^2} \rho _{14}(t)-i\frac{A_n}{\hbar}\rho _{14}(t)\nonumber\\
   &-\frac{i \left(\Omega \rho _{11}(t)-\Omega \rho _{44}(t)+2 \epsilon \rho _{14}(t)\right)}{\hbar}, \label{eq:main}
\end{align}
where $A_n=\sigma_n\left[(E-\epsilon)^{n/2}-(E+\epsilon)^{n/2}\right]^2$
quantifies the attenuation of the off-diagonal component $\rho_{14}$ and effectively determines the decoherence timescale of the model. The parameter $n$ exclusively influences the constant $A_n$, meaning that studying the qualitative behavior of this Lindblad evolution only requires specifying initial conditions and values for $A_n$.

The decoherence timescale in our approach is given by $\tau_{D,n}=A_n^{-2}\hbar^2/t_{QG}$, or equivalently (recalling that $\ell=p_P^{-1}=c/E_P$):

\begin{equation}\label{eq:dec-time2}
\tau_{D,n}=\frac{2^{2-n}}{\left[(E-\epsilon)^{n/2}-(E+\epsilon)^{n/2}\right]^2}\frac{\hbar^2}{(Mc^2)^{4-n}}\frac{E_P^2}{t_{QG}}.
\end{equation}

\begin{figure}[htbp]
    \centering
    \begin{tabular}{cc}
        \includegraphics[width=0.48\linewidth]{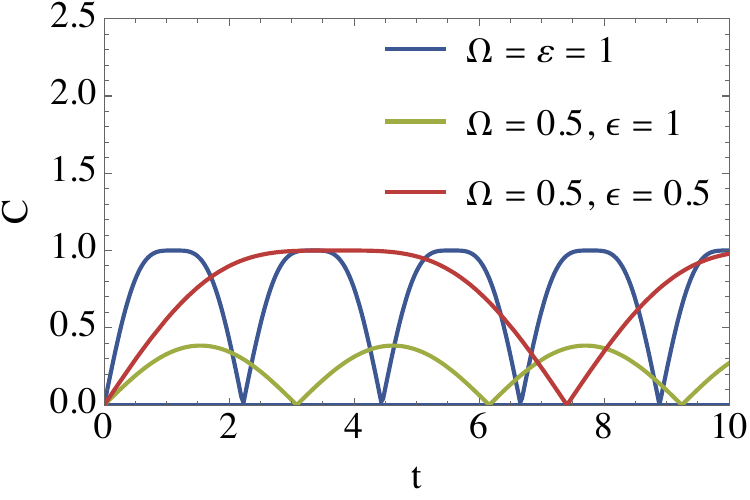} & 
        \includegraphics[width=0.48\linewidth]{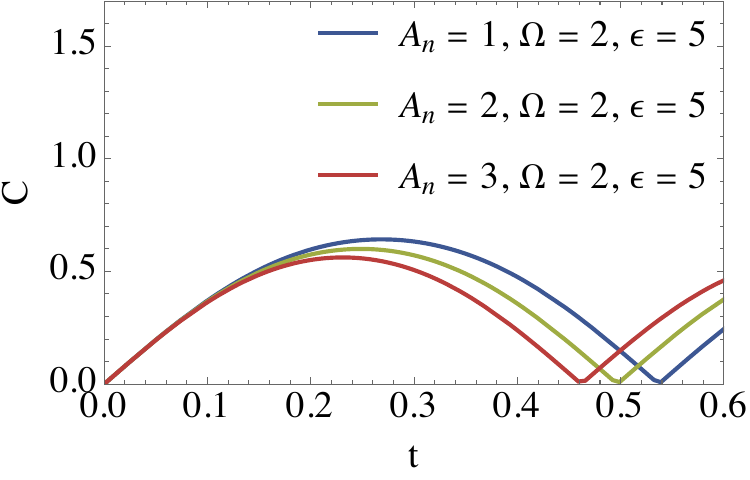} \\
        (a) Without LIV & (b) Systematic LIV \\ \\
        \includegraphics[width=0.48\linewidth]{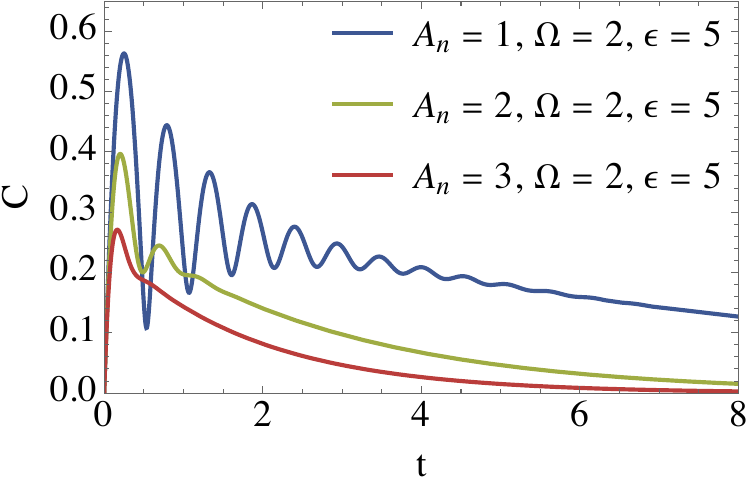} & 
        \includegraphics[width=0.48\linewidth]{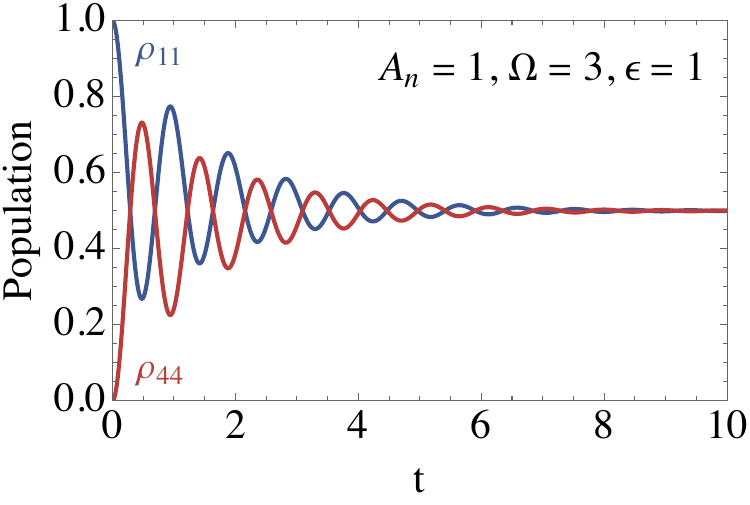} \\
        (c) Stochastic LIV & (d) Stochastic LIV
    \end{tabular}
    \caption{Figs.(a)-(c) describe the concurrence for an initially unentangled state ($\theta=0$) for the cases without LIV ($A_n=0$), with systematic LIV ($t_{QG}=0$), and with stochastic LIV ($t_{QG}=1$), respectively. Fig.(d) describes the population with stochastic LIV ($t_{QG}=1$). We set $\hbar=c=1$.}
    \label{fig:grid}
\end{figure}

In Fig.(\ref{fig:grid}\textcolor{red}{b}), we analyzed the systematic LIV case, i.e., in which the quantum gravity scale do not fluctuate, which is achieved by requiring $t_{QG}=0$ in \eqref{eq:main}. The larger is the LIV parameter, the more attenuated is the maximum of the concurrence. This means that even in the absence of fluctuations of the quantum gravity scale in the dispersion relation, the LIV parameter suppresses entanglement formation. From \eqref{eq:main}, we see that the solution is given by Eq.\eqref{eq:sol1}-\eqref{eq:sol2}, where we map $\epsilon\mapsto \epsilon+A_n/2$

For the stochastic case, i.e., when $t_P\neq 0$, Fig.(\ref{fig:grid}\textcolor{red}{c}) demonstrates that the interaction initially dominates the system's evolution, generating entanglement between particles. However, LIV effects gradually dampen the density matrix's off-diagonal elements. The oscillation extrema remain largely unchanged as they are governed by the entanglement timescale \eqref{eq:ent-time}, which depends solely on $\Omega$ and $\epsilon$. The characteristic timescales depend on the dispersion relation details encoded in parameter $n$ through \eqref{eq:dec-time2}.

Figure (\ref{fig:grid}\textcolor{red}{d}) shows the population dynamics for $\theta=0$ (initially separable state). When $\Omega>\epsilon$, the stronger interaction drives pronounced oscillations between $\rho_{11}$ and $\rho_{44}$ before decoherence effects ultimately equilibrate them to 0.5.

\subsection{Mesoscopic Particles}

Following \cite{Bose:2017nin}, we consider particles with separation $d=200\,\mu\text{m}$, position shift $L=100\,\mu\text{m}$, mass $M=10^{-14}\,\text{kg}$, and frequency $\omega=\epsilon/\hbar=1\,\text{Hz}$. The resulting entanglement timescale $\tau_E\sim 1\,\text{s}$ matches the requirements for measurable gravitationally induced entanglement phases in the proposed experimental setup. While these numerical values coincide, they represent distinct physical quantities. A shorter $\tau_E$ would simply compress the oscillations shown in Fig.(\ref{fig:grid}\textcolor{red}{c}), with entanglement remaining observable on average if decoherence is sufficiently suppressed.

For these parameters, the LIV decoherence timescale becomes $\tau_{D,1}\approx 2\times10^{19}\,\text{s}\sim 600$ billion years for $n=1$, increasing further for larger $n$. This represents a lower bound, as arguments suggesting the quantum gravity energy scale should grow with particle number would push this timescale even higher. Consequently, LIV effects would not produce measurable decoherence in this configuration.

To assess the generality of this conclusion, we note that \cite{Bose:2017nin} predicts measurable gravitational entanglement between masses $m_1$ and $m_2$ over timescale $\tau$ when:

\begin{equation}
    \frac{G m_1m_2}{\hbar (d-L)}\tau+ \frac{G m_1m_2}{\hbar (d+L)}\tau\sim 1.
\end{equation}

For the reasonable $\tau\sim 1$s timescale, this requires $Gm_1m_2/D\sim \hbar\,\text{s}$, where $D$ is the system length scale. Meanwhile, the LIV decoherence timescale becomes $\frac{2^{2-n}\hbar}{(Mc^2)^{4-n}E^n}E_P^3$ when approximating $(\Delta E^{n/2})^2\sim E^n$ and setting $t_{QG}=t_P=\hbar/E_P$. For sufficiently high energy scales $E$, this fundamental LIV decoherence could compete with coherence maintenance requirements:

\begin{equation}
\frac{2^{2-n}\hbar}{(Mc^2)^{4-n}E^n}E_P^3\sim 1\,\text{s}.
\end{equation}

This corresponds to energy scales: $n=1$: $E_{n=1}\approx 2.5\times10^{-15}\,\text{J}\approx 15.6\,\text{keV}$ ($\omega_1\sim10^{18}\,\text{Hz}$); $n=2$: $E_{n=2}\approx 10^{-6}\,\text{J}\approx 6.2\,\text{TeV}$ ($\omega_2\sim10^{27}\,\text{Hz}$); $n=3$: $E_{n=3}\approx 0.8\,\text{mJ}\approx 5\,\text{PeV}$ ($\omega_3\sim10^{30}\,\text{Hz}$).
%\begin{itemize}
 %   \item $n=1$: $E_{n=1}\approx 2.5\times10^{-15}\,\text{J}\approx 15.6\,\text{keV}$ ($\omega\sim10^{18}\,\text{Hz}$)
  %  \item $n=2$: $E_{n=2}\approx 10^{-6}\,\text{J}\approx 6.2\,\text{TeV}$ ($\omega\sim10^{27}\,\text{Hz}$)
   % \item $n=3$: $E_{n=3}\approx 0.8\,\text{mJ}\approx 5\,\text{PeV}$ ($\omega\sim10^{30}\,\text{Hz}$)
%\end{itemize}

This implies in the need for a relativistic treatment beyond our Galilean framework for practical reasons. For the considered mesoscopic system, collisional and thermal decoherence sources dominate over LIV effects \cite{Rijavec:2020qxd}.

\subsection{Generalization to Different Lindbladians}

The parameter $n$, which determines the correction power, appears exclusively in the decoherence factor $A_n$ (Eq.~\eqref{eq:main}). This allows generalization to cases involving different momentum operator powers $\hat{K}^{n/2}$ and alternative $\sigma_n$ values. For example, a generic Lindblad equation:

\begin{equation}\label{eq:gen-main}
    \partial_t\rho(t)=-\frac{i}{\hbar}\left[\hat{H}_0,\rho(t)\right]-\sigma_n\left[\hat{K}^{n/2}(t),\left[\hat{K}^{n/2}(t),\rho(t)\right]\right],
\end{equation}
with $n=4$ and $\sigma_n=\sigma_{PI}=16 M^2\ell_P^4 t_P/\hbar^6$, recovers the Petruzziello-Illuminati model based on the Generalized Uncertainty Principle \cite{Petruzziello:2020wkd}. The decoherence time $\tau_D^{PI}=1/(\sigma_{PI}(\Delta E^2)^2)$ becomes $\tau_D^{PI}\sim10^{141}$ s for our mesoscopic particles. Reference \cite{Donadi:2024amp} analyzed similar decoherence effects in mechanical oscillators.

%%%%%%%%%%%%%%%%%%%%%%%%%%%%%%%%%%%%%%%%%%%%%%%%%%%%%%%%%%%%%%%%%%%%%%%%%%%%%%%%%%%%%%%%%%%%%%%%%%%%%%%%%%%%%%%%%%%%

\section{Final Remarks}\label{sec:6}

Using tools presented in \cite{Petruzziello:2020wkd}, we constructed a Lindblad equation based on a modified dispersion relation with a fluctuating Planck-scale parameter. In this framework, the quantum spacetime -- characterized by deviations from the standard relativistic expression - acts as an unavoidable environment that necessarily interacts with every quantum system. Equipped with this evolution equation, we analyzed its impact on a system composed of two cat states interacting gravitationally in a superposition of position eigenstates (gravcat states). For closed quantum systems, these particles evolve into an entangled pair mediated by Newtonian gravitational interaction. We calculated the concurrence as an entanglement quantifier for this system and observed its oscillatory behavior over time.

When incorporating terms originating from quantum spacetime effects, we examined the system's evolution and found that entanglement gradually vanish due to the influence of stochastic Planck-scale parameters. Our analysis revealed that because quantum gravity effects are extremely small, the timescale for such fluctuations to dissipate quantum correlations is remarkably long. Consequently, such systems would more likely experience decoherence from other environmental factors before quantum spacetime effects become significant in the infrared regime. (The ultraviolet regime appears to bring this timescale to more accessible levels, though this would require more sophisticated analytical tools.)

The gravcat states we considered represent a simplified version of systems discussed in the literature concerning interactions between spatial superposition states. Therefore, it would be valuable for future investigations to examine more detailed proposals, such as those in \cite{Bose:2017nin,Marletto:2017kzi,Krisnanda:2019glc,Rijavec:2020qxd}, before drawing definitive conclusions. This is particularly important because quantum gravity effects, even in the infrared regime, can sometimes produce unexpected amplifications of apparently mild effects \cite{Amelino-Camelia:2009wvc}.

%%%%%%%%%%%%%%%%%%%%%%%%%%%%%%%%%%%%%%%%%%%%%%%%%%%%%%%%%%%%%%%%%%%%%%%%%%%%%%%%%%%%%%%%%%%%%%%%%%%%%%%%%%%%%%%%%%%%%%%%%%%%%%%%%%%%%%%%%%%%%%%%%%%%%%%%%%%%%%%%%%%%%%%%%%%%%%%%%%%%%%%%%%%%%%%%%%%%%%%%%%%%%%%%%%%%%%%%%%%%%%%%%%%%%%%%

\section*{Acknowledgments}
I. P. L. was partially supported by the National Council for Scientific and Technological Development - CNPq,
grant 312547/2023-4. G. V. thanks Coordena\c c\~ao de Aperfei\c coamento de Pessoal de Nível Superior - Brazil (CAPES) - Finance Code 001 and the National Council for Scientific and Technological Development - CNPq,
grant 140335/2022-6 for financial support.  
V. B. B. was partially supported by the National Council for Scientific and Technological Development- CNPq, grant No. 307211/2020 7. M. R. acknowledges FAPEMIG Grant No. APQ 02226/22. M. R.
was partially supported by the National Council for Scientific and Technological Development - CNPq,
grant 311565/2025-5. K.S is funded through an undergraduate scholarship by the Universidade Federal da Paraíba (UFPB) - PIBIC program. The authors would like to acknowledge networking support by the COST Action BridgeQG (CA23130) and the COST Action RQI (CA23115), supported by COST (European Cooperation in Science and Technology).

\bibliographystyle{utphys}
\bibliography{mdr-ent}

\end{document}